\documentclass[iicol]{sn-jnl}

\usepackage{graphicx}%
\usepackage{array}
\usepackage[caption=false,font=normalsize,labelfont=sf,textfont=sf]{subfig}
\usepackage{stfloats}
\usepackage{url}
\usepackage{verbatim}
\usepackage{siunitx}
\usepackage{CJKutf8}
\usepackage{caption}
\usepackage{cuted}
\usepackage{balance}
\usepackage{lineno}  
\usepackage{soul} 
\soulregister\cite7
\soulregister\ref7 
\soulregister\SI7

\usepackage{multirow}%
\usepackage{amsmath,amssymb,amsfonts}%
\usepackage{amsthm}%
\usepackage{mathrsfs}%
\usepackage[title]{appendix}%
\usepackage{xcolor}%
\usepackage{textcomp}%
\usepackage{manyfoot}%
\usepackage{booktabs}%
\usepackage{algorithm}%
\usepackage{algorithmicx}%
\usepackage{algpseudocode}%
\usepackage{listings}%

\usepackage{tikz,xcolor}

\hypersetup{hidelinks, colorlinks=true, allcolors=black, pdfstartview=Fit, breaklinks=true}
\definecolor{lime}{HTML}{A6CE39}
\DeclareRobustCommand{\orcidicon}{
    \begin{tikzpicture}
        \draw[lime, fill=lime] (0,0)
        circle[radius=0.16]
        node[white]{{\fontfamily{qag}\selectfont \tiny \.{I}D}};
    \end{tikzpicture}
    \hspace{-2mm}
}
\foreach \x in {A, ..., Z}{%
    \expandafter\xdef\csname orcid\x\endcsname{\noexpand\href{https://orcid.org/\csname orcidauthor\x\endcsname}{\noexpand\orcidicon}}
}

\theoremstyle{thmstyleone}%
\theoremstyle{thmstyletwo}%

\theoremstyle{thmstylethree}%

\graphicspath{{./figure/}}  
\begin{document}


\title[Spin-NeuroMem: A Low-Power Neuromorphic Associative Memory Design Based on Spintronic Devices]{Spin-NeuroMem: A Low-Power Neuromorphic Associative Memory Design Based on Spintronic Devices}


\author[1]{\fnm{Siqing} \sur{Fu}\hspace{-1.5mm}\orcidA{}}\email{fusiqingnudt@nudt.edu.cn}

\author[1]{\fnm{Lizhou} \sur{Wu}\hspace{-1.5mm}\orcidC{}}\email{lizhou.wu@nudt.edu.cn}

\author*[1]{\fnm{Tiejun} \sur{Li}}\email{tjli@nudt.edu.cn}

\author[1]{\fnm{Chunyuan} \sur{Zhang}}\email{cyzhang@nudt.edu.cn}

\author[1]{\fnm{Jianmin} \sur{Zhang}}\email{jmzhang@nudt.edu.cn}

\author[1]{\fnm{Sheng} \sur{Ma}}\email{masheng@nudt.edu.cn}

\affil[1]{\orgdiv{College of Computer Science and Technology}, \orgname{National University of Defense Technology}, \orgaddress{\street{Deya Road}, \city{Changsha}, \postcode{410073},  \country{China}}}


\abstract{Biologically-inspired computing models have made significant progress in recent years, but the conventional von Neumann architecture is inefficient for the large-scale matrix operations and massive parallelism required by these models. This paper presents Spin-NeuroMem, a low-power circuit design of Hopfield network for the function of associative memory. Spin-NeuroMem is equipped with energy-efficient spintronic synapses which utilize magnetic tunnel junctions (MTJs) to store weight matrices of multiple associative memories. The proposed synapse design achieves as low as 17.4\% power consumption compared to  the state-of-the-art synapse designs. Spin-NeuroMem also encompasses a novel voltage converter with a 53.3\% reduction in transistor usage  for effective Hopfield network computation. In addition, we propose an associative memory simulator for the first time, 
which achieves a 5M$\times$ speedup with a comparable associative memory effect.
By harnessing the potential of spintronic devices, this work paves the way for the development of energy-efficient and scalable neuromorphic computing systems.}

\keywords{Neuromorphic computing, Associative memory, Spintronic devices, Low-power}



\maketitle

\section{Introduction}
\label{Sec:Introduction}

Neuromorphic computing (NC) \cite{Amrouch2021, markovic2020physics} mimics brain functionalities through complex connections between a large number of artificial neurons and synapses, resulting in powerful computing capabilities. Owning to its great potential for applying to energy-efficient pattern recognition, associative memory, and decision-making beyond the traditional von Neumann architecture, NC has become a strong candidate to evolve into a new computing paradigm in the future. The goal of NC research is to emulate neurons and synapses of the human brain by capturing the behaviors of emerging devices at nanoscale, overcoming the limitations of traditional computing modes. As a typical feedback-based NC model, Hopfield network maps input patterns to stable output states to achieve various functionalities including associative memory, error correction, categorization, familiarity recognition, and time sequence retention \cite{hopfield1986computing}. Among these functionalities, associative memory is the most promising application of Hopfield networks, attracting great research attention\cite{hu2015associative} due to its ability to restore the complete picture of a given data set from partial information, similar to human memory.


Efficient execution of NC relies on the prerequisite of hardware implementation. Conventionally, hardware implementations of the Hopfield network are typically based on CMOS technology, which faces challenges related to area and power consumption. In recent years, the emergence of new devices such as memristors \cite{li2018review} offers an opportunity. However, NC systems demand repeated current stimulation to memristive synapses, leading to device resistance drift. This inevitably instigates weight variations that damage the reliability of synapse \cite{Telminov2022}. Additionally, many challenges on endurance and defect rates need to be addressed when using memristors. Unlike memristors, spintronic devices such as magnetic tunnel junctions (MTJs) provide new possibilities for reliable synaptic design thanks to the fact that they exploit electron spin rather than electron charge for memory read and write \cite{Zhang2016, fukami2018perspective, amirany2020nonvolatile}. However, designing advanced spintronic-based NC systems still faces many challenges, including: 1) the production of special  MTJs remains difficult \cite{Zhang2016}; 2) insufficient device reliability under process variations (PVs) \cite{fukami2018perspective}; 3) dramatic increase in power consumption as the number of synaptic weights  increases \cite{amirany2020nonvolatile}. Therefore, it is imperative to design a reliable neural computing system with scalable synaptic weights, while achieving low power consumption and high PV tolerance.

In this paper, we present a low-power neuromorphic associative memory design named Spin-NeuroMem. It utilizes spintronic devices to design synapses for storing weight matrices for multiple associative memories. The proposed synapse design significantly reduces power consumption compared to existing solutions. 
The non-volatile property of MTJs allows our circuit to be completely powered off during inactive phases, which further reduces the leakage power of our design. 

Our contributions in this paper can be summarized as follows:

\begin{itemize}
\item We present a novel voltage converter for hardware-based Hopfield networks. Our design utilizes a modified logic gate circuit to obtain  binary-to-Hopfield-network conversion, resulting in a 53.3\% reduction in transistor count compared to the existing work.
\item We propose a spintronic synapse composed of MTJ matrices that can provide different weights to support neural computation. Our design is remarkably energy-efficient, with a power consumption of only 17.4\% of the previous work for ten positive-weight synapses.
\item We develop an associative memory simulator to evaluate the performance of our Spin-NeuroMem at large scale. By evaluating the performance of the simulated Spin-NeuroMem, we demonstrate its nearly equivalent associative memory effect compared to the software-based Hopfield network, achieving a 5.05e6$\times$ speedup.
\end{itemize}

The structure of this paper is organized as follows. In Section~\ref{Sec:Background}, we review the basic principles of associative memory and the fundamental concepts of MTJ technology. Section~\ref{Sec:Design} provides a detailed explanation of the design principles and circuit implementation of Spin-NeuroMem. In Section~\ref{Sec:Experiments}, we evaluate and analyze the associative memory functionality of Spin-NeuroMem at the circuit and system levels. Finally, Section~\ref{Sec:Conclusion} concludes the entire paper.

\section{Background}
\label{Sec:Background}

\subsection{Hopfield Network and Associative Memory}

\begin{figure}[t]
\centering 
\includegraphics[width=7cm]{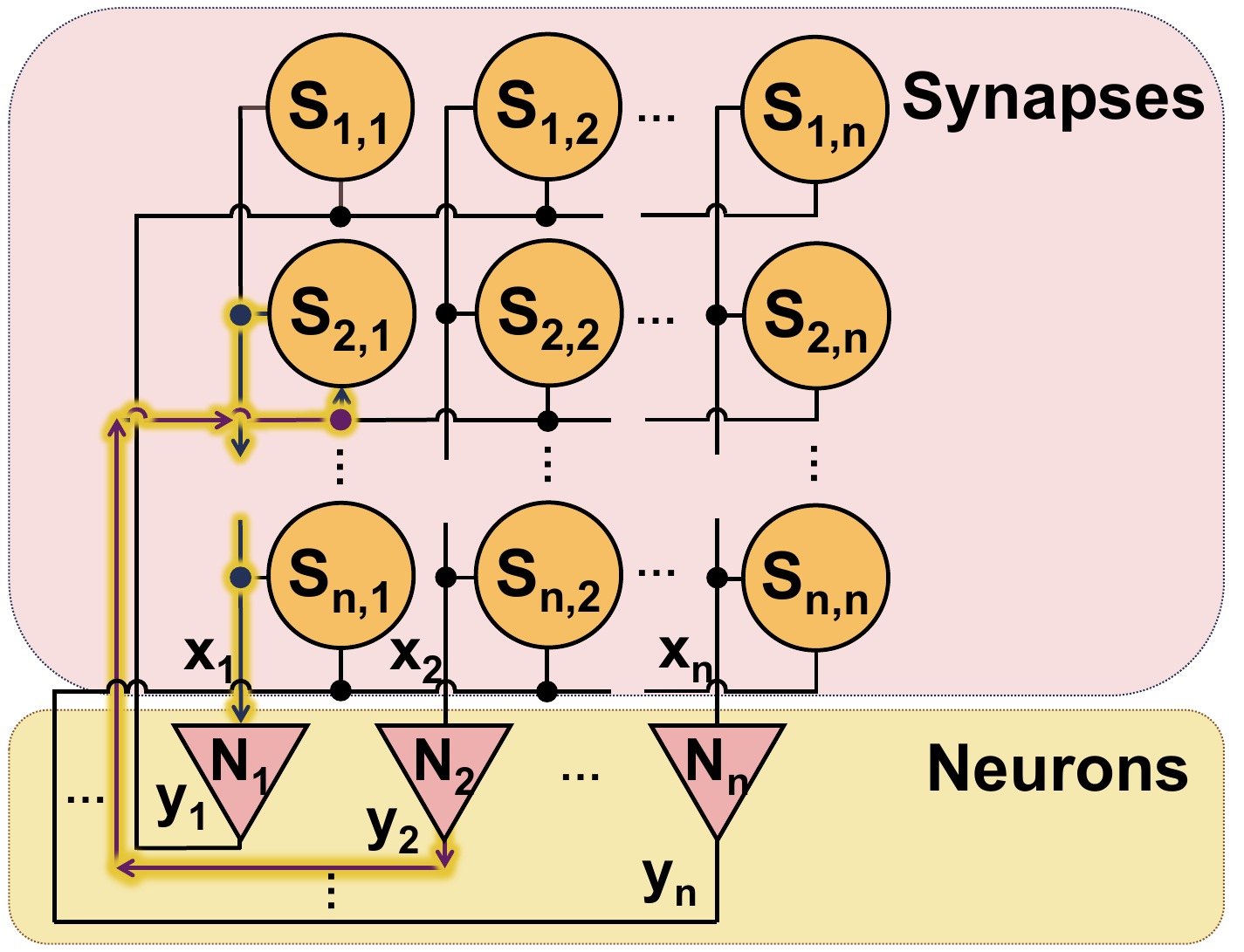}
\caption{The structure of Hopfield network with $n$ dimensions.}
\label{Figure1_topology}
\end{figure}

The Hopfield neural network \cite{hopfield1986computing} is generally used to solve combinatorial optimization problems or implement associative memory for pattern recognition. Associative memory is similar to the human brain memory that can recall the memorized data by providing a portion of the data or noisy data rather than by giving an address in the existing semiconductor memories \cite{hu2015associative}.

The Hopfield network is a single-layer, fully connected recurrent neural network composed of $n$ neurons and $n^{2}$ synapses, as shown in Fig.~\ref{Figure1_topology}. The working principle of the Hopfield network model can be expressed by:


\begin{align}
x_{j}(t+1)&=\sum_{i=1}^{n}w_{i,j}\times    y_{i}(t),   \quad x_j, y_i\in\{-1,1\}, \label{eq:hopfield_eq1}\\
y_{j}(t+1)&=f(x_{j}(t+1)), \label{eq:hopfield_eq2}\\
f(x)&=\left\{\begin{matrix}1 \,\,\, , \quad \, x\ge \theta_{j}  
 \\-1 \, , \quad \,  x<  \theta_{j}  .
\end{matrix}\right. \label{eq:hopfield_eq3}
\end{align}

In the above equations,  $w_{i,j}$ represents the weight of synaptic $S_{i,j}$ connecting the $i$-th and $j$-th neurons $N_{i}$ and $N_{j}$, $y_i(t)$ represents the output of the $i$-th neuron at time $t$. $x_j(t+1)$ represents the state of the $j$-th neuron at time $t+1$; it is calculated by summing up every row of the product $w_{i,j}\cdot y_{i}$ along column $j$ at time $t$.  The output $y_j(t+1)$ of neuron at $t+1$  is determined by the function $f$ and $x_j(t+1)$.  $\theta_j$ represents the threshold of the $j$-th neuron. For instance, as the highlighted path in Fig.~\ref{Figure1_topology} illustrates, when the presynaptic neuron $N_{2}$ outputs $y_2(t)$ at time $t$, it is transmitted through the synaptic $S_{2,1}$ to the postsynaptic neuron $N_{1}$. The electrical potential $x_1(t+1)$, which accumulates all the incoming signals to $N_{1}$, determines whether or not $N_{1}$ is activated, thus concluding a round of neural signal transmission.

To memorize $m$ patterns, each of which is denoted as a vector $P_{k}=\left( a_{1},a_{2},\dots,a_{n}  \right) $, the learned result of the weight matrix $W$ can be derived as:

\begin{equation}
W=\sum_{k=0}^{m}P_{k}\times  P_{k}^{T}.    
\label{equ4}
\end{equation}

Note that each element of the matrix $w_{i,j}\in\{-m, -m+1, \ldots, m\}$.

\subsection{Magnetic Tunnel Junction}

MTJs are widely used spintronic devices that have a three-layer structure \cite{Fu2023, Gallagher,chen2012tunable,chen2012yoke}. As shown in Fig.~\ref{Figure2_mtj}, this structure consists of two ferromagnetic layers separated by a dielectric tunnel barrier (TB) layer. The lower ferromagnetic layer, referred to as the pinned layer (PL), has its magnetization fixed along the easy axis of the MTJ\cite{Ikeda2007}. The upper ferromagnetic layer, referred to as the free layer (FL), can have its magnetization parallel (P) or antiparallel (AP) to that of the PL\cite{yuasa2018materials}. Due to the tunnelling magneto-resistance (TMR) effect\cite{mathon2001theory}, the resistance value ($R_{AP}$) is higher in the AP state and referred to as logic ``1'', while in the P state, the resistance value ($R_{P}$) is lower and referred to as logic ``0''. The difference between these two resistance values is expressed by the TMR ratio:

\begin{figure}[t]
\centering
\includegraphics[width=7cm]{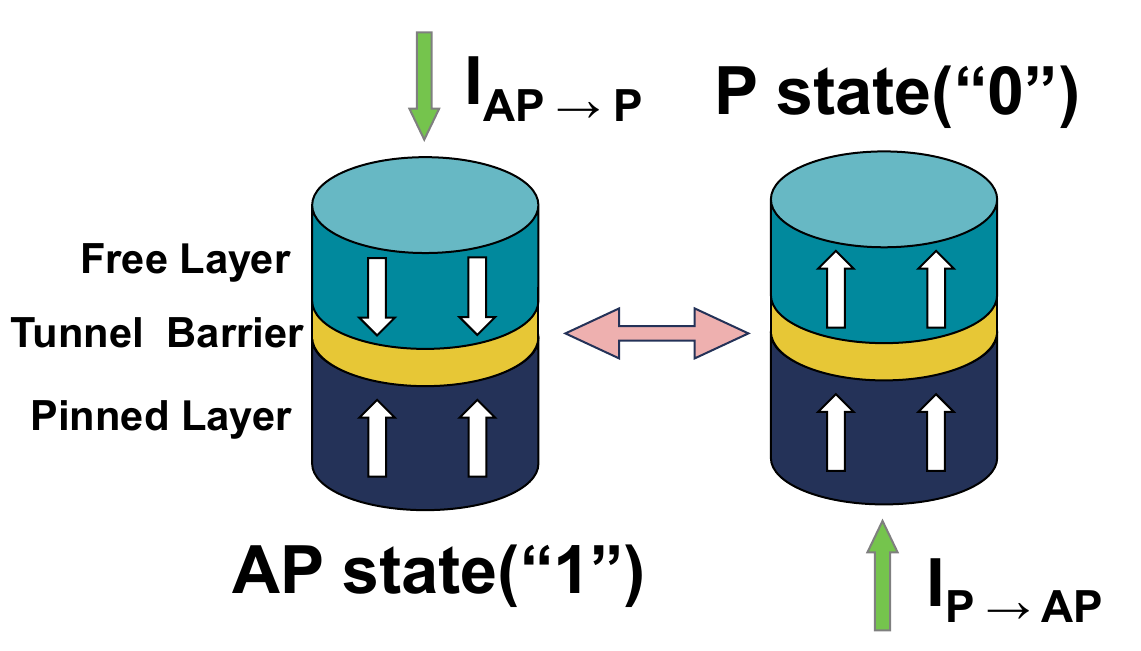}
\caption{The MTJ structure and STT-based write mechanism.} 
\label{Figure2_mtj}
\end{figure}

\begin{equation}
\mathrm{TMR}=\frac{(R_{AP}-R_{P})}{R_{P}}\times100\%.
\label{equ5}
\end{equation}

The resistive state of MTJ can be switched by applying a spin-polarized current \cite{Sato2020VLSI}. Fig.~\ref{Figure2_mtj} shows that a positive pulse across the MTJ in the AP state drives a current $I_{\mathrm{AP}\rightarrow\mathrm{P}}$ perpendicularly across from the FL to the PL. When certain thresholds for pulse amplitude and width are surpassed (typically $2$-\SI{100}{\nano\second}), the magnetization of the FL switches direction. In a similar manner, a negative pulse exceeding the critical switching current under the spin-polarized current $I_{\mathrm{P}\rightarrow\mathrm{AP}}$ can switch the MTJ from P to AP. Due to the stable binary magnetic states (i.e., AP and P), $R_{\mathrm{AP}}$ and $R_{\mathrm{P}}$ do not show a degradation trend as found in memristors over $10^7$ writing cycles. \cite{Carboni2018}.

In summary, MTJs are perfect candidates for synaptic design, owning to non-volatility, re-programability, low-power. In addition, MTJs feature almost no resistance drift over time, which  overcomes the limitations of hardware NC systems based on memristors.

\subsection{Related Work}
Next, we review the research advancement in NC implementation based on novel devices, including memristors and spintronic devices.
\subsubsection{Memristor-based Neuromorphic Hardware}

The work in \cite{hu2015associative} demonstrates the implementation of associative memory using a memristive Hopfield network. It presents adjustable resistance in memristors for pattern storage and retrieval, as well as programmable synaptic weights in a 3-bit memristive Hopfield network. The design in \cite{cai2020power} adopts a memristor-based annealing system with a neuromorphic architecture, providing a high-throughput solution for NP-hard problems through parallel operations and leveraging hardware noise for improved efficiency.

Despite some pioneering attempts, the application of memristors in NC is limited by its physical characteristics. For example, the resistive drift over time caused by electric field changes and atomic migration inevitably leads to variations in synaptic weights \cite{Ebong2011}. Additionally, many challenges on durability and defect rates need to be addressed when using memristors \cite{Ebong2011, likharev2008hybrid, Saadeldeen2013}.

\subsubsection{Spintronic-based Neuromorphic Hardware}

Unlike memristors, spintronic devices such as MTJs provide new possibilities for reliable synaptic design thanks to the fact that they exploit electron spin rather than electron charge for memory read and write.

The compound spintronic synapse design in \cite{Zhang2016} shows promise for NC with its stable multiple resistance states, but challenges remain in addressing PVs and achieving consistent material and thickness of stacked MTJs. The spintronic synapse proposed in \cite{fukami2018perspective} demonstrates associative memory operations using an antiferromagnet/ferromagnet heterostructure driven by spin-orbit torque, but its stability against process, voltage, and temperature variations remains challenging due to device variability and non-linearity.

Owing to the low-power and non-volatile characteristics of spintronic devices, recent research \cite{Amirany2019,amirany2020nonvolatile,Nasab2022,Rezaei2022,Nasab2023,Rezaei2023,Rezaei2024a,Rezaei2024b,Rezaei2024c} has focused on the design of associative memory using spintronic devices. In these works, MTJs provide configurability, nonvolatility, and high endurance to the design, while CNTFETs compensate for the limitations of conventional transistors in deep nanoscale nodes.

The work of Rezaei et al. \cite{Rezaei2023} aims to increase synaptic capacity by utilizing parallel-connected MTJs to form synapses. The studies by Amirany et al. \cite{amirany2020nonvolatile} and Rezaei et al. \cite{Rezaei2024c}, as representatives, provide synapses with multi-weight storage for associative memory through a series-connected MTJ design. Although the design offers significant power advantages over its CMOS counterpart, the synaptic design, which uses serially-connected MTJs for multiple weights, inevitably leads to increased power consumption. Moreover, the voltage adder required for each synapse occupies unnecessary on-chip area.


\section{Proposed Spin-NeuroMem Design}
\label{Sec:Design} 

In this section, we first provide an overview of  the proposed Spin-NeuroMem design. Thereafter, we elaborate the structures and functionalities of each component in the design.

\subsection{Design Overview}

\begin{figure}[t]
\centering
\includegraphics[width=0.5\textwidth]{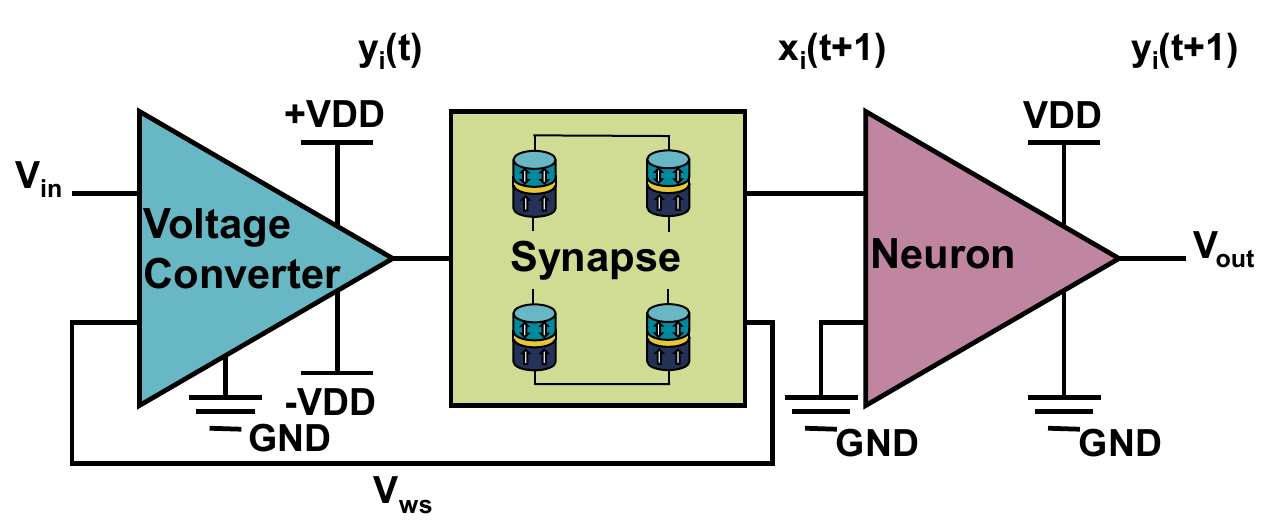}

\caption{Proposed Spin-NeuroMem design with three elemental components: voltage converter, synapse, and neuron.} 
\label{Figure3_Spin-NeuroMem}
\end{figure}

Fig.~\ref{Figure3_Spin-NeuroMem} shows constituent parts of Spin-NeuroMem, including voltage converters, synapses, and neurons. The voltage converter takes a binary value (0 or 1) as input from external or feedback from presynaptic neurons and outputs a bipolar value (-1 or 1) for synaptic activation. The synaptic activation generates an analog voltage that contains weight information, which is then transmitted to postsynaptic neurons. The postsynaptic neuron receives incoming signals from all connected presynaptic neurons through synapses, sums them up, and updates its output value through an activation function, which finally results in a binary value (0 and 1). This process corresponds to a neural activation from a presynaptic neuron to a postsynaptic neuron, as highlighted in Fig.~\ref{Figure1_topology}.

\begin{figure}[t]
\centering
\includegraphics[width=6cm]{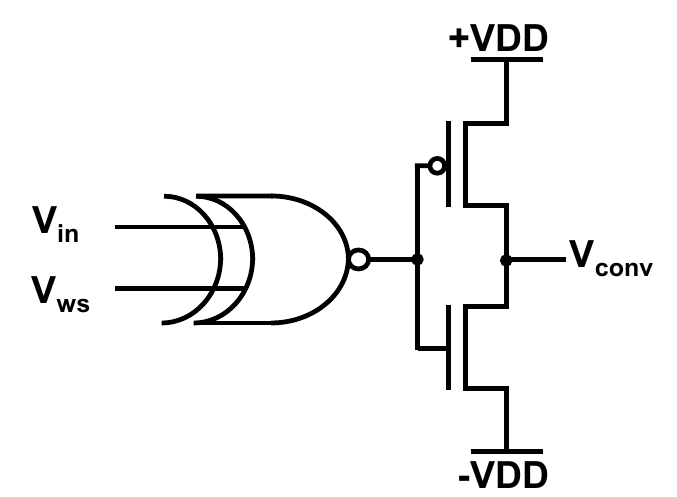}

\caption{Proposed voltage converter design.} 
\label{Figure4_VC}
\end{figure}

\subsection{Voltage Converter Design}

\begin{table}[t]
\caption{Binary-to-bipolar logic conversion table.}\label{tab1}
\begin{tabular}{@{}ccc@{}}
\toprule
\textbf{$V_{\mathrm{in}}$}     & \textbf{$V_{\mathrm{ws}}$} & \textbf{$V_{\mathrm{conv}}$}  \\
\midrule
0 & 0 & -1  \\ 
0 & 1 & 1  \\ 
1 & 0 & 1  \\ 
1 & 1 & -1  \\
\botrule
\end{tabular}
\end{table}

The voltage converter design includes an XNOR gate and a modified inverter-like structure, as shown in Fig.~\ref{Figure4_VC}. The XNOR gate takes inputs $V_{\mathrm{in}}$ and $V_{\mathrm{ws}}$, representing the input from the presynaptic neuron and the sign of the weight read from the synapse, respectively. The output of the voltage converter, $V_{\mathrm{conv}}$, is the converted output voltage that is then provided to the synapse. Table~\ref{tab1} presents the logic values of $V_{\mathrm{in}}$, $V_{\mathrm{ws}}$, and $V_{\mathrm{conv}}$ along with their conversion relationships.

\begin{figure}[t]
\centering
\includegraphics[width=0.48\textwidth]{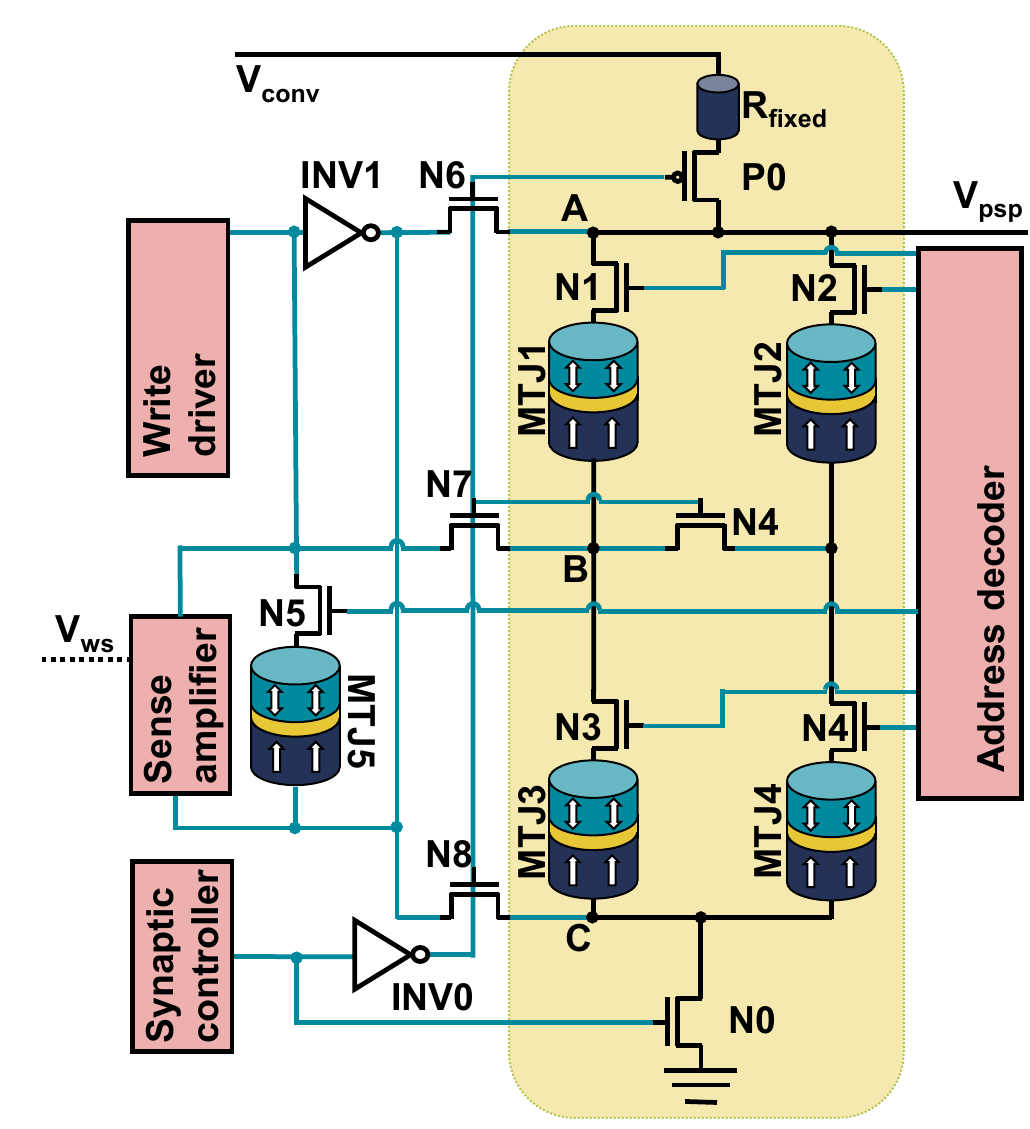}

\caption{Spintronic synapse design which is a non-volatile memory and computational unit composed of a full MTJ array.} 
\label{Figure5_Sy}
\end{figure}

The proposed voltage converter significantly saves on-chip area compared to the prior  CNTFET-based voltage adder\cite{amirany2020nonvolatile}. 
The evaluation of  area employs the same methodology as that used in \cite{YAN2021}. For the \SI{45}{\nano\meter} technology node, the XNOR gate in the proposed voltage converter comprises 6 NMOS transistors and 6 PMOS transistors. The total number of NMOS and PMOS transistors is 7 each. In the layout, the area of the NMOS transistors is \SI{0.149}{\square\micro\meter}, and the PMOS transistors have an area of \SI{0.214}{\square\micro\meter}. Therefore, the total area of the proposed voltage converter is \SI{2.542}{\square\micro\meter}. For the single voltage adder proposed in \cite{amirany2020nonvolatile}, the MUX is composed of one OR gate, two AND gates, and one NOT gate, totaling 10 NMOS and 10 PMOS transistors. Including an additional 5 NMOS and 5 PMOS transistors in the rest of the circuit, the total area is \SI{5.448}{\square\micro\meter}. This results in a 53.3\% reduction in area. In other words, implementing a Hopfield network capable of processing the MNIST dataset \cite{lecun1998mnist}, composed of 784 neurons and 614656 synapses, could save an area of \SI{1.786}{\milli\meter^2} approximately.

\subsection{Synapse Design}

In the information transmission process, neurotransmitters are released by pre-synaptic neurons and can affect the action potential of post-synaptic neurons via the synapses. Our spintronic synapses have been designed to mimic this communication process, providing varying weights as depicted in Fig.~\ref{Figure5_Sy}. Each synapse comprises $N\times N+1$ MTJs ($N=2$ in this case), including $N\times N$ MTJs for controlling the weight values and one MTJ for controlling the weight sign. This results in a total of $N\times N+1$ positive weights and $N\times N+1$ negative weights.

Our synapse design can work in two different modes, i.e., associative memory mode and configuration mode, depending on the signal from the synamptic controller. When the synaptic controller outputs ``1'', the associative memory mode is activated. In this case, transistors N0 and P0 are turned on, while transistors N4, N6, N7, and N8 are turned off. Focusing on the black wire section of the circuit, we observe that each of the four MTJs has different resistance values in the AP and P states due to the TMR effect. Consequently, five weight configurations determine the synaptic strength: 4$R_{AP}$, 3$R_{AP}$1$R_{P}$, 2$R_{AP}$2$R_{P}$, 1$R_{AP}$3$R_{P}$, and 4$R_{P}$. The input of the synapse is $V_{\mathrm{conv}}$ corresponds to the voltage converter output, and the output is the postsynaptic potential voltage ($V_{\mathrm{psp}}$), which will be transmitted to the postsynaptic neuron. Assuming $R1$, $R2$, $R3$ and $R4$ are the resistance values of the four MTJs in the $2\times 2$ MTJ matrix, and $R_{\mathrm{fixed}}$ is the fixed resistance, $V_{\mathrm{psp}}$ can be expressed as:
\begin{equation}
\resizebox{0.4\textwidth}{!}{$
V_{\mathrm{psp}} = \frac{(R1+R3)(R2+R4)}{R_{\mathrm{fixed}} \left( \sum_{i = 1}^{4} R_i \right) + (R1+R3)(R2+R4)} V_{\mathrm{conv}}
$}
\label{equ6}
\end{equation}
In order to achieve the maximum swing of $V_{\mathrm{psp}}$, the value of $R_{\mathrm{fixed}}$ should be approximately halfway between $R_{P}$ and $R_{AP}$. Note that some weight configurations, like 2$R_{AP}$2$R_{P}$, correspond to different MTJ matrix configurations (e.g., $R1$ and $R2$ or $R1$ and $R3$ configured as AP). However, we only program one of them as the effective weight to ensure large and uniform weight differences. MTJ5 in Fig.~\ref{Figure5_Sy} memories the weight sign, which is read out by the sense amplifier and fed back to the voltage converter to control $V_{\mathrm{conv}}$ direction achieving fully non-volatile storage of the weight values.

When the synaptic controller outputs ``0'', the configuration mode is activated. MTJs receive write current from bottom to top or top to bottom depending on the output of the write driver. The address decoder controls the gate of transistors N1, N2, N3, N4, and N5 that are connected in series with the MTJ. They are turned on to select the MTJ to be configured. A more detailed description of the configuration process can be found in Section \ref{subsec:circuit_sim}. It is worth noting that the synapse configuration cost is not a concern as weight rewriting occurs only once during the process of  weight learning.

Our 5-MTJ synapse design can be extended for more weight requirements as the total resistance range of spintronic synapses remains unchanged. A perpendicular MTJ based on the MgO/CoFeB structure has achieved a TMR of 249\% \cite{wang2018current}. Our design builds on the current achievable advanced manufacturing process. Higher TMR ratio in  MTJs of the future will allow for more weighting options in spintronic synapses.

\subsection{Neuron Design}

\begin{figure}[t]
\centering
\includegraphics[width=0.4\textwidth]{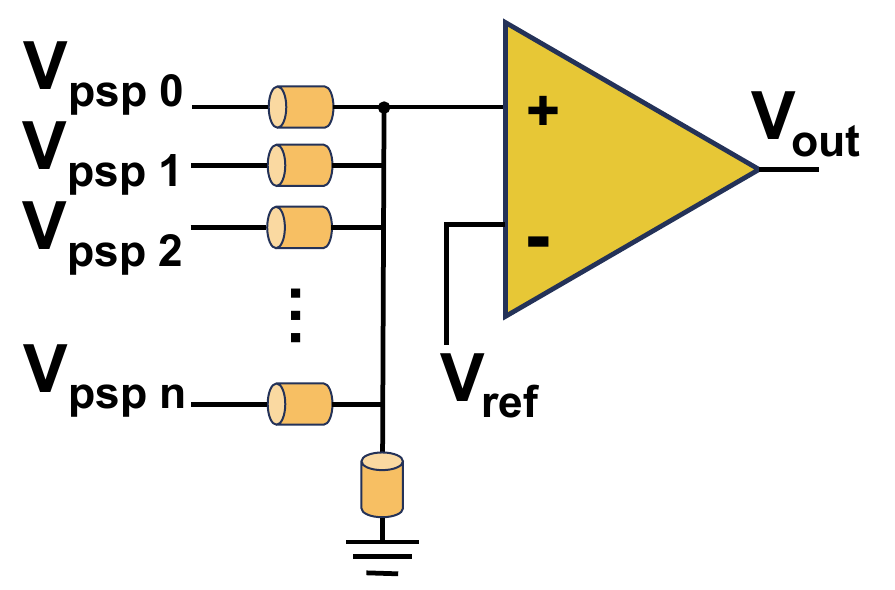}

\caption{Neuron design in Spin-NeuroMem.} 
\label{Figure6_Ne}

\end{figure}

The neuron design originates from the CNTFET neuron proposed in \cite{amirany2020nonvolatile}. In Fig.~\ref{Figure6_Ne}, the $N$ presynaptic neurons output postsynaptic potentials through synapses. After calculating $\sum V_{\mathrm{psp}} $, the resistive voltage adder within the neuron transmits the result to a single pin of a CMOS-based comparator. The other pin is the reference voltage $V_{\mathrm{ref}}$, which is set to \SI{0}{\volt}. Once the sum of the voltages exceeds the threshold, the neuron is activated and outputs ``1''; otherwise, it remains inactive and outputs ``0''. 

\section{Experiments and Evaluation}
\label{Sec:Experiments}

In this section, we first elaborate the experimental setups at both circuit and system levels. Thereafter, we present circuit simulation results of Spin-NeuroMem and evaluate its functionalities, performance, and power consumption.
In addition, we perform system-level experiments and evaluation using an in-house Python simulator. To demonstrate the advantage of our proposed design, we also compare the performance of Spin-NeuroMem with that of the prior work as well as software implementations of associative memory.



\begin{table*}[t]
\centering
\caption{Key device parameters for MTJ compact model.}\label{tab2}
\resizebox{0.65\textwidth}{!}{
\begin{tabular}{c c c}
\toprule
{ \textbf{Parameter}} & {\textbf{Description}}  & {\textbf{Value}}  \\ \midrule
{$t_{\mathrm{FL}}$}       & {Thickness of the free layer}        & {\SI{1.3}{\nano\meter}}  \\ 
{$\sigma _{t_{\mathrm{FL}}} $}       & {Standard deviation of $t_{\mathrm{FL}}$}        & {3\% of \SI{1.3}{\nano\meter}}  \\

{$CD$}         & {Critical diameter}        & {\SI{32}{\nano\meter}}   \\ 
{$t_{\mathrm{TB}}$}       & {Thickness of the tunnel barrier}     & {\SI{0.85}{\nano\meter}} \\ 

{$\sigma _{t_{\mathrm{TB}}} $}       & {Standard deviation of $t_{\mathrm{TB}}$}     & {3\% of \SI{0.85}{\nano\meter}} \\ 

{$\mathrm{TMR}$}       & {TMR ratio} & {249\%}   \\ 
{$\sigma _{\mathrm{TMR}}$}       & {Standard deviation of TMR} & {3\% of 249\%}   \\ \botrule
\end{tabular}}
\end{table*}

\subsection{Experimental Setup}

\begin{figure}[t]
    \begin{minipage}[t]{\linewidth}
        \centering
        \includegraphics[width=1\textwidth]{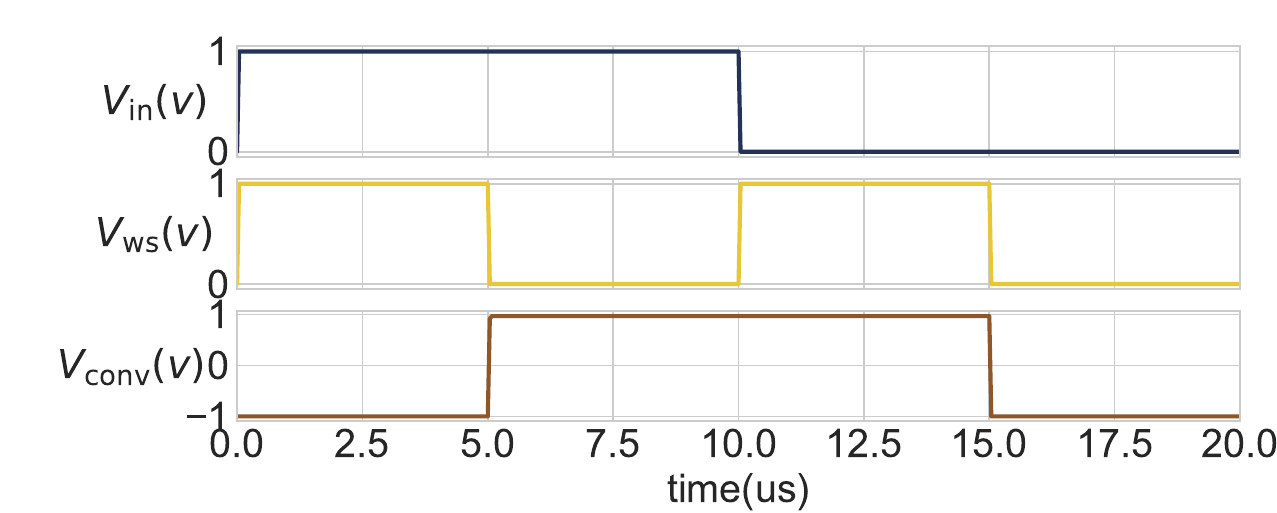}
        \centerline{(a) Voltage converter}
     \
    \end{minipage}%
    \\
    \begin{minipage}[t]{\linewidth}
        \centering
        \includegraphics[width=1\textwidth]{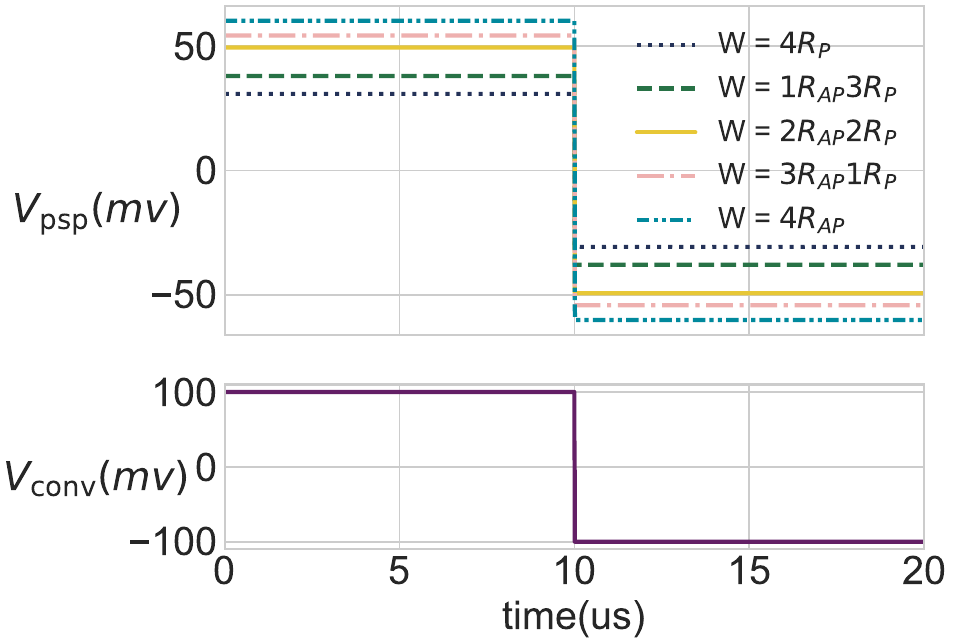}
        \centerline{(b) Synapse}
 \
    \end{minipage}%
    \\
    \caption{Transient simulation of voltage converter ansynapse and neuron in Spin-NeuroMem.}
\label{Figure7_sim}

\end{figure}

We conducted circuit simulations using Cadence Virtuoso tools with the  MTJ compact model in \cite{Wu2022} and GPDK \SI{45}{\nano\meter} technology. We took into account PV and estimated synapse weight drifts through Monte Carlo simulations. The critical parameters of the MTJ model and its PV strengths are provided in Table~\ref{tab2}. The TMR value is consistent with the current capabilities of advanced manufacturing processes \cite{wang2018current}.  Note that PV is introduced by considering 3$\sigma$ deviation for the key device parameters.
All circuit-level simulations were conducted under the ambient temperature of \SI{300}{\kelvin}. Additionally, to facilitate a fair comparison, the previous work was re-conducted with identical parameters.

Due to the exponential overhead in time and computing resources to simulate a large-scale associative memory, circuit simulation is unsuitable to evaluate the performance of Spin-NeuroMem at the system level.
Consequently, we have developed a Python-based simulator, which will be open-sourced. To ensure simulation accuracy and consistency, circuit parameters were extracted from comprehensive circuit simulations and subsequently fed into the simulator. This ensures our simulator accurately replicates the circuit functionalities and performance exhibited during circuit-level simulations. 
The software-based Hopfield network was developed using Python 3.7 in both serial and parallel modes. The code was run on Ubuntu 20.04.1 with an Intel i9-12900 CPU. We compared the system-level performance using two metrics which were recall rate and recall latency.

\begin{figure}[t]
\centering
\includegraphics[width=0.5\textwidth]{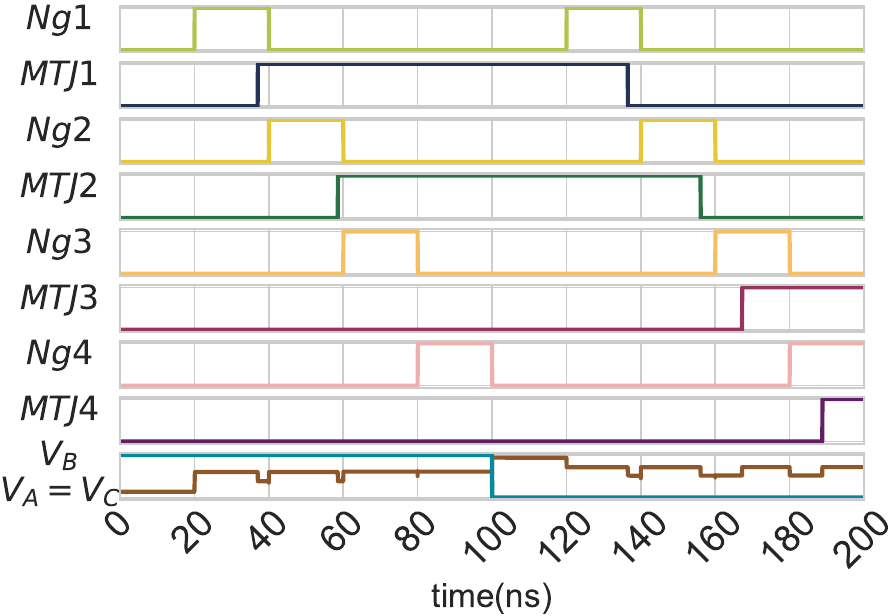}

\caption{Transient simulation  of  synaptic weight configuration.} 
\label{Figure8_Wr}

\end{figure}

\begin{figure}[t]
\centering
\includegraphics[width=0.5\textwidth]{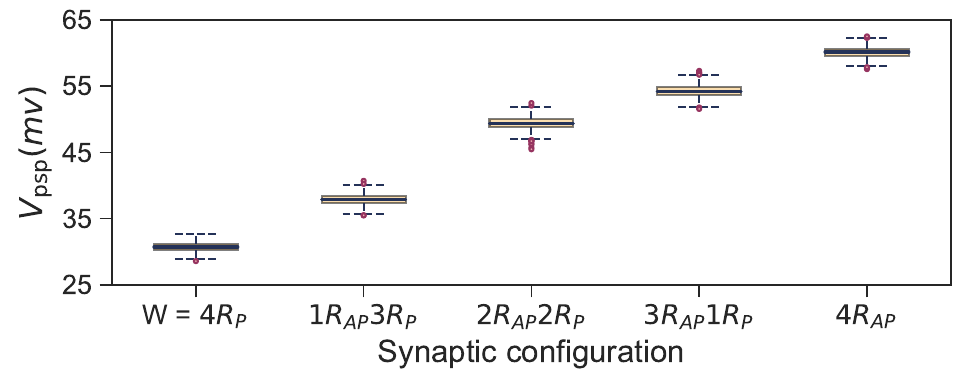}

\caption{Monte Carlo simulation results of output voltage of spin neuronal synapses under process variations.} 
\label{Figure9_MC}

\end{figure}

\subsection{Circuit Simulation}
\label{subsec:circuit_sim}

\subsubsection{Functional evaluation}

Fig.~\ref{Figure7_sim}(a) depicts the functionality of the voltage converter via transient simulation.  It can be seen that a $V_{\mathrm{in}}$ of ``1'' results in a $V_{\mathrm{conv}}$ of ``1'' if the synaptic weight sign ($V_{\mathrm{ws}}$) is positive, otherwise it would be ``-1''. Similarly, when $V_{\mathrm{in}}$ is ``0'' and $V_{\mathrm{ws}}$ is positive, the resultant $V_{\mathrm{conv}}$ value is ``-1''; otherwise, it would be ``1''. The complete binary-to-bipolar conversion relations can be found in Table~\ref{tab1}.

Fig.~\ref{Figure7_sim}(b) shows the diverse weight selection capabilities of the all-spin neural synapse. Note that $V_{\mathrm{conv}}$ is transmitted from the previous stage. When it is ``1'', five positive weight outputs are generated depending on different MTJ resistance configurations of the $2\times2$ MTJ array. In a similar manner, five negative weight outputs are produced when $V_{\mathrm{conv}}$ is ``-1''.



Fig.~\ref{Figure8_Wr} presents the functional simulation of the synaptic weight configuration process, when the synaptic controller output is set to ``0'' (configuration mode). In the figure, Ng1-Ng4 correspond to the gate signals of the NMOS transistors that select the four MTJ devices (N1-N4) shown in Fig.~\ref{Figure5_Sy}, while MTJ1-MTJ4 denote the magnetization states of the corresponding MTJ devices; $V_{A}$, $V_{B}$, and $V_{C}$ represent voltages at points A, B, and C, respectively.

The write driver initially outputs a high signal for \SI{100}{\nano\second}. When the NMOS transistor connected in series with the MTJ is turned on at this time, the MTJ array can receive write current from both A-B and C-B directions. In the initial state of the simulation, all four MTJs are in P-state, representing a logic ``0''. The gate voltages of N1, N2, N3, and N4 increase sequentially by \SI{20}{\nano\second}. MTJ1 and MTJ2 are written to ``1'', while MTJ3 and MTJ4 are configured to ``0''. Subsequently, the write driver outputs a low signal for \SI{100}{\nano\second}. If the NMOS transistor connected in series with the MTJ is turned on, the write current flows through the MTJ array in the opposite direction. After a delay, the four MTJs are set to ``0'', ``0'', ``1'', ``1'', respectively.

\subsubsection{Impact of device variations on weight}

To evaluate the functionality of spin-based synapses in the presence of PV, we took into account a 3\% variation in the parameters listed in Table~\ref{tab2} in the MTJ model and conducted Monte Carlo simulations. We conducted 1$\,$000 simulations for each synaptic weight configuration, resulting in a  total of 5$\,$000 simulations considering only positive weights in synaptic connections based on a $2\times 2$ MTJ matrix neural synapse shown in Fig.~\ref{Figure5_Sy}. In Fig.~\ref{Figure9_MC}, we observed that the upper and lower quartiles of output voltage, obtained through the synaptic weights, showed a significant difference for the $2\times 2$ MTJ matrix-based neural synapse.
\begin{figure}[b]
\centering
\includegraphics[width=0.5\textwidth]{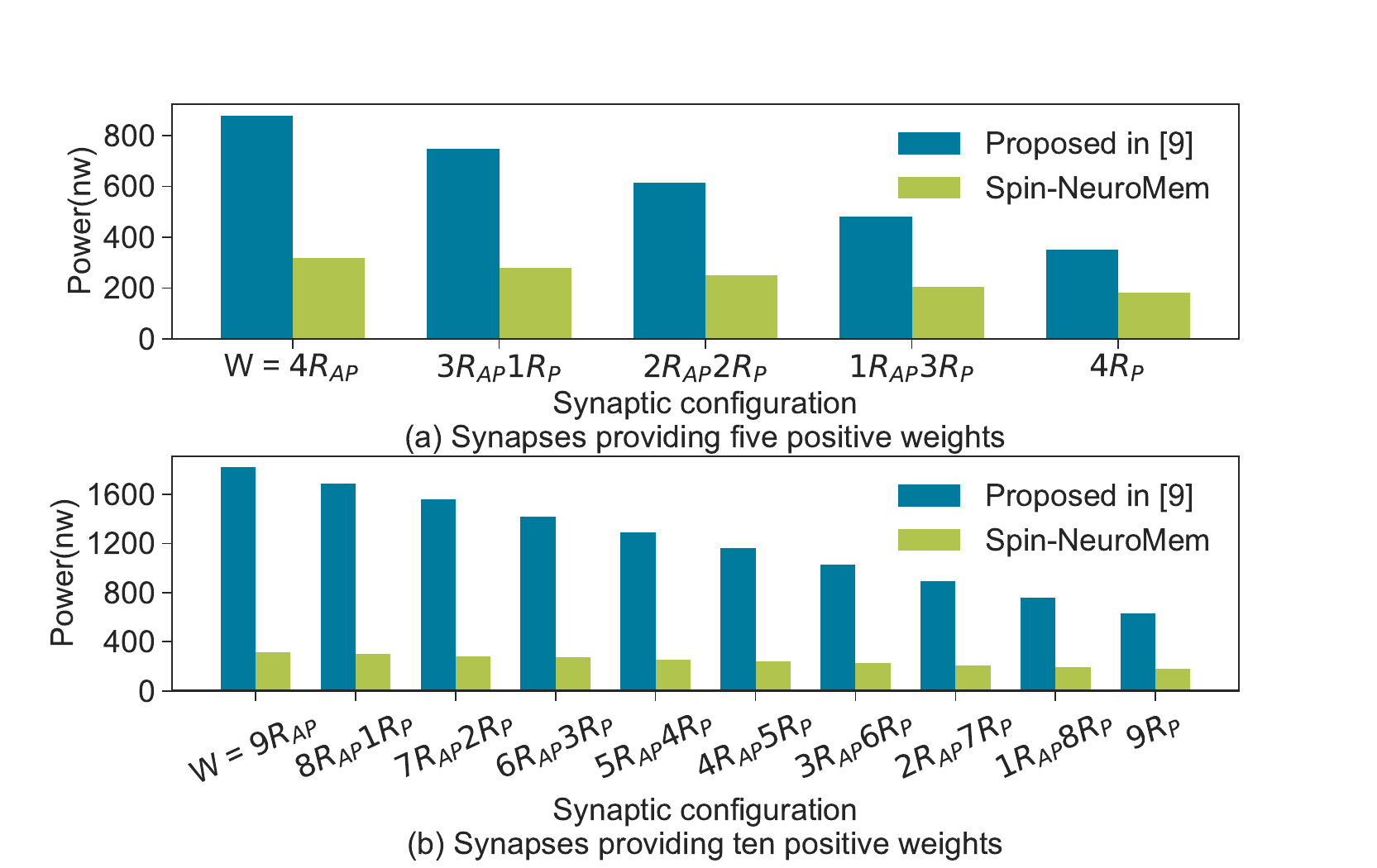}

\caption{Power consumption comparison between proposed spin synapse and \cite{amirany2020nonvolatile} at different scales.} 
\label{Figure10_Pw}

\end{figure}

\subsubsection{Power consumption}

To compare the power consumption of our proposed design with  previous  work, we conducted a comparison between the power consumption of the synaptic connection presented in this paper and that in \cite{amirany2020nonvolatile}, under the same transistor and MTJ process parameters. Fig.~\ref{Figure10_Pw}(a) shows five neural synapses based on $2\times 2$ MTJ matrices, providing five positive weights. Our design significantly reduces power consumption, ranging from 36.1\% to 32.2\% of the previous work under the five synaptic weights, measured in $\mathrm{mW}$. Furthermore, our design exhibits minimal increase in power consumption as the number of weights increases. Fig.~\ref{Figure10_Pw}(b) shows a comparison  between the power consumption of synapses that use more MTJs to provide more weights. Increasing the scale of reconfigurable MTJs in the spintronic synapse results in a noticeable increase in power consumption in \cite{amirany2020nonvolatile}. while our design maintains the same power consumption range, ranging from 17.4\% to 28.9\% of the previous work, measured in $\mathrm{mW}$.

It is important to note that the average power consumption of a single synapse is not affected by the network size; rather, the power consumption of a synapse is influenced by the size of the MTJ array constituting the synapse. For Hopfield networks, the network memory capacity depends on synaptic precision, which requires a larger MTJ array in each synapse for multi-value storage. As shown in Fig.~\ref{Figure10_Pw}, the previously proposed series-connected MTJ arrays\cite{amirany2020nonvolatile,Rezaei2024c} will experience a significant increase in power consumption in this scenario, while the synaptic power consumption of Spin-NeuroMem does not rise with the increase in MTJ array size.

\subsection{Systematic Performance Evaluation}

To evaluate the effectiveness of the proposed design in processing associative memory tasks, we conducted  systematic experiments using a constructed Hopfield network shown in Fig.~\ref{Figure1_topology}. We created two Hopfield networks of different scales: 1) 100 neurons and 10$\,$000 synaptic connections for processing binary matrices of 10$\times$10 pixels, and 2)  784 neurons and 614$\,$656 synapses for processing the MNIST dataset which has binary matrices of 28$\times$28 pixels. 



\begin{figure}[t]
    \begin{minipage}[t]{0.5815\linewidth}
        \centering
        \includegraphics[width=\textwidth]{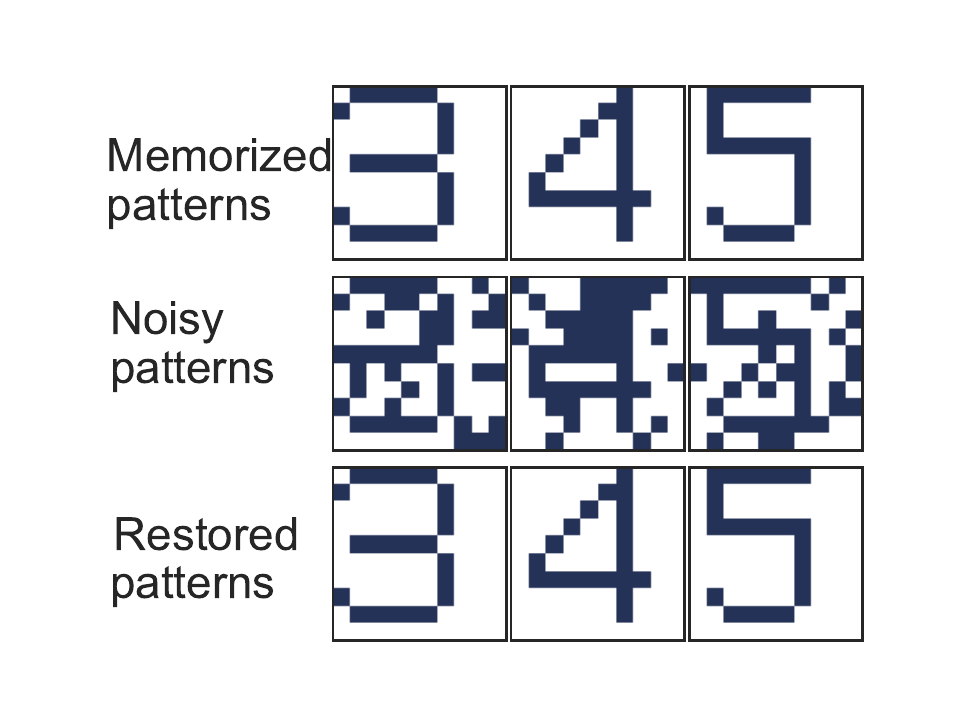} 
        \hfill 
        \centerline{\quad\quad\quad\quad\quad(a)10x10 pixel digits.}
    \end{minipage}%
    \begin{minipage}[t]{0.4185\linewidth}
        \centering
        \includegraphics[width=\textwidth]{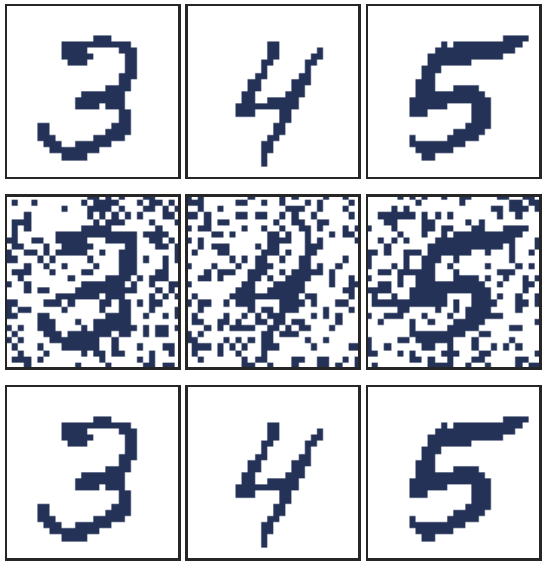}
        \centerline{(b)MNIST dataset.}
    \end{minipage}
   
    \caption{Demonstration of successful associative memory recall with noisy input patterns by two sizes of Hopfield networks.}
    \label{Figure11_pattern}
    
\end{figure}

Multiple input patterns with local similarities and well-distributed patterns are employed to evaluate the effect of multiple associative recalls. Fig.~\ref{Figure11_pattern}(a) shows a successful recovery of 100-dimensional pattern vectors which are randomly injected with noise. The memorized patterns, input patterns with 30\% noise, and recovered patterns after associative recall are shown separately in this figure. Fig.~\ref{Figure11_pattern}(b) utilizes a relatively larger-scale network to process the MNIST dataset and demonstrates the ability to recover noisy data effectively.

\begin{figure}[t]
\centering
\includegraphics[width=0.45\textwidth]{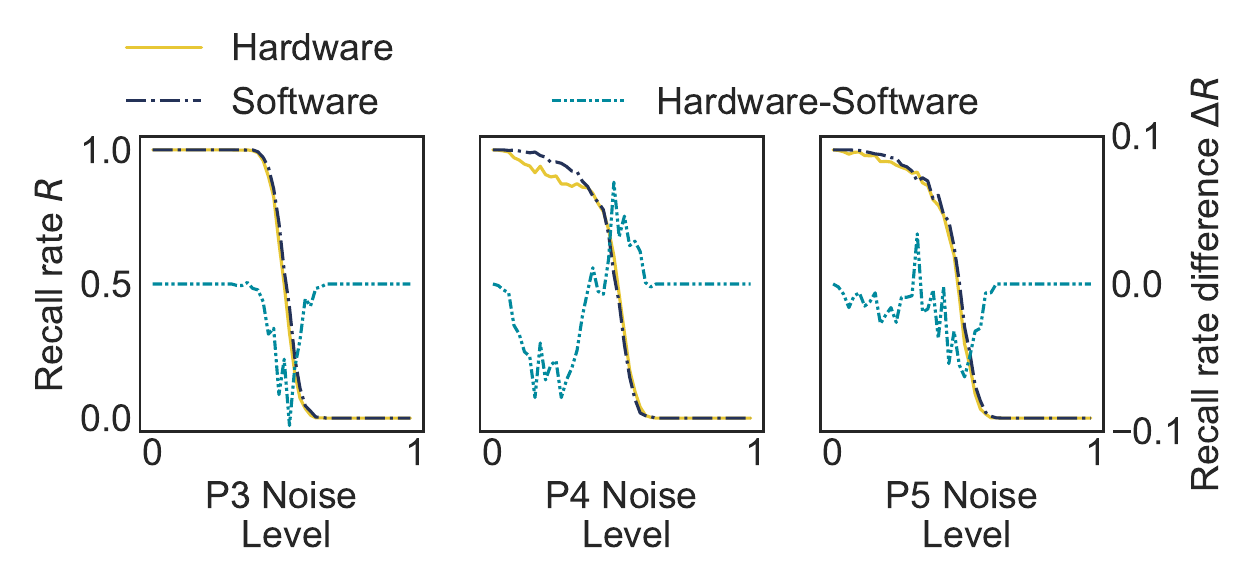}

\caption{Comparison of recall rates for noise pattern restoration of length 100 using Spin-NeuroMem and software Hopfield network.} 
\label{Figure12_sys}

\end{figure}

\begin{figure}[t]
\centering
\includegraphics[width=0.48\textwidth]{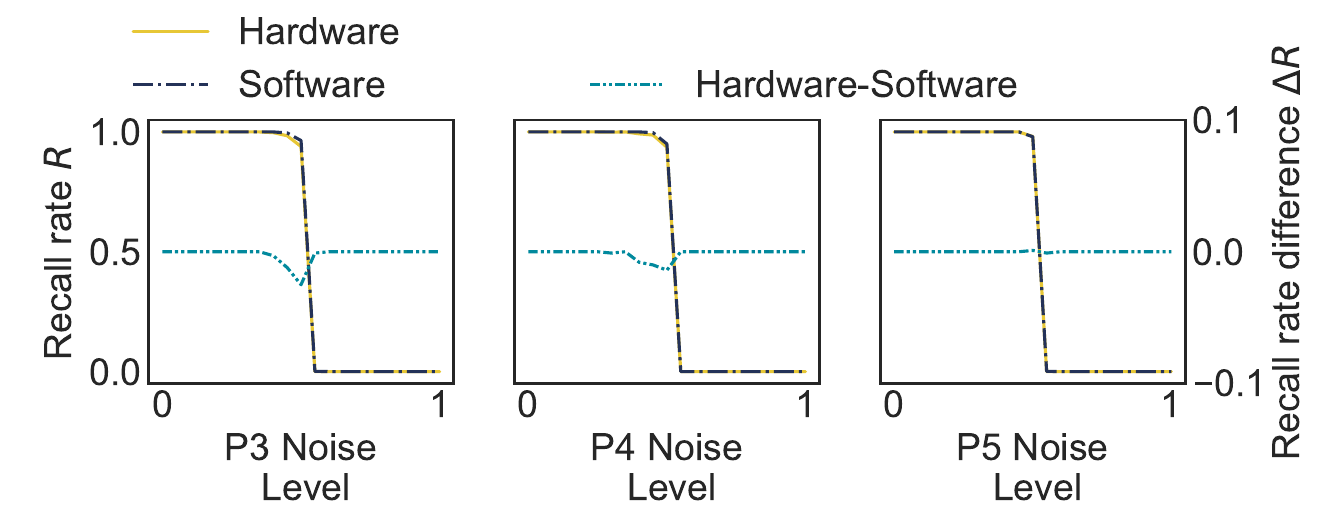}
\caption{Comparison of recall rates for noise pattern restoration of MNIST dataset using hardware and software network implementations.} 
\vspace*{-1.1\baselineskip} 
\label{Figure13_sysnew}
\end{figure}

\begin{table*}[t]
\centering
\caption{Comparison of recall latency between Spin-NeuroMem and software-based Hopfield networks.}
\label{tab3}
\resizebox{0.7\textwidth}{!}{
\begin{tabular}{c c c c} 
\toprule
 &\textbf{  single-core CPU }     & \textbf{   multi-core (24) CPU } & \textbf{Spin-NeuroMem}    \\  \midrule

recall latency (s)&$5.5 \times 10^{-3}$&$5.6 \times 10^{-4}$&$1.09 \times 10^{-9}$  \\ 
speedup &$1$&$9.82$&$5.05 \times 10^{6}$  \\ \bottomrule

\end{tabular}
}

\end{table*}

Fig.~\ref{Figure12_sys}  shows the recall rate $R$ for patterns ``3'', ``4'', and ``5'' (denoted as P3, P4, P5) with a size of $ 10 \times 10$ pixels, under different noise levels. The network executed associative recall on input noisy patterns  1$\,$000 times for each noise level to calculate $R$ value. Different colored curves represent the recall rates $R$ and their variations under software and hardware implementations. The secondary y-axis represents the difference in recall rates between the two implementations $(\Delta R=R(\mathrm{hardware-software}))$. It can be observed that around a noise level of 50\%, the recall rate of the hardware implementation is slightly lower than that of the software implementation due to possible errors introduced by representing weights using post-synaptic voltage. We performed further analysis on the significant difference between the hardware and software implementations using the Mann-Whitney U test\cite{macfarland2016mann}, with the alternative hypothesis being that the median of the second sample is greater than the median of the first sample. The calculated p-value is 0.33, which is greater than 0.05, and thus, we cannot reject the null hypothesis. 
In other words, the recall effect of Spin-NeuroMem is comparable to the software implementation. 

Fig.~\ref{Figure13_sysnew} presents a similar conclusion drawn from restoring MNIST digits. Due to its larger scale, the associative memory process of this network exhibits a greater degree of fault tolerance. Through $20 \times 1\,000$ recall trials at 5\% intervals from 0\% to 100\% noise rates, the results still support the previous experimental analysis. The expansion of Hopfield networks offers the advantage of broader synaptic connections, leading to more stable recall rate evaluations across varying noise levels and reduced discrepancies between hardware and software. 

Due to the characteristics of Hopfield networks, when the noise rate exceeds a threshold, pixel correlations are disrupted, reducing recall rate. This applies to both software implementations and Spin-NeuroMem, which should avoid high noise rate scenarios.

Table~\ref{tab3} compares the recall latency for a single recall task using Spin-NeuroMem and software-based Hopfield networks. The input patterns of the serialized and paralyzed networks use a same noisy 28$\times$28 pixel MNIST image. The CPU execution time is derived from the runtime statistics of the software running on a real CPU, while the execution time for Spin-NeuroMem comes from SPICE simulation. The execution time for a single software associative memory recall  is \SI{5.5}{\milli\second} on average when utilizing a single CPU core. The CPU accelerates computations through multi-core parallel processing. 
A 24-core CPU achieves a $9.82\times$ speedup compared to a single-core CPU. In contrast, the novel and efficient architecture of Spin-NeuroMem exhibits a  gate-level latency of \SI{1086}{\pico\second}. It achieves a  speedup of $5.05\times10^{6}$ in associative memory recall compared to its software counterpart running on a single-core CPU. Due to the lack of layout for the MTJ model, parasitic capacitances have been neglected, resulting in an overestimation of the performance of Spin-NeuroMem. Nevertheless, the results still effectively highlight the performance advantages of hardware neuromorphic architectures in executing Hopfield networks.

\vspace*{-0.1\baselineskip} 
\section{Conclusion}
\label{Sec:Conclusion}

This paper presents Spin-NeuroMem, a low-power neuromorphic associative memory design that integrates spintronic devices and CMOS components. The experimental results show superior performance of this design in terms of both power consumption and area, particularly as the weight scale increases. Moreover, our proposed Spin-NeuroMem can achieve a recall rate on par with that of software-based Hopfield networks while showcasing a significant improvement in speed. Overall, our work demonstrates the potential of spintronic neural network hardware for building  next-generation  neural computing platforms.

\bibliographystyle{sn-aps}
\bibliography{ref}

\balance
\end{document}